# Magnetic resonance in iron oxide nanoparticles: quantum features and effect of size

Maxim M.Noginov[1], N. Noginova[2], O. Amponsah[2], R. Bah[2], R. Rakhimov[2], and V. A. Atsarkin[3]

1. Applied Physics, Cornell University, Ithaca, NY
2. Norfolk State University, Norfolk, VA
3. Institute of Radio Engineering & Electronics RAS, Moscow, Russia

In order to better understand the transition from quantum to classical behavior in spin system, electron magnetic resonance (EMR) is studied in suspensions of superparamagnetic magnetite nanoparticles with an average diameter of ~ 9 nm and analyzed in comparison with the results obtained in the maghemite particles of smaller size (~ 5 nm). It is shown that both types of particles demonstrate common EMR behavior, including special features such as the temperature-dependent narrow spectral component and multiple-quantum transitions. These features are common for small quantum systems and not expected in classical case. The relative intensity of these signals rapidly decreases with cooling or increase of particle size, marking gradual transition to the classical FMR behavior.

**Introduction**

Magnetic nanoparticles are promising systems with a wide range of applications in industry, science, and medicine, such as data storage, MRI contrast agents, medical drug delivery and hyperthermia (for a review, see [1]). Furthermore, they present fundamental physics problems associated with magnetization dynamics on the nanoscale.

Ferromagnetic domains containing many coupled spins are commonly considered as classical systems while the quantum approach is necessary to describe behavior when only one or several coupled spins are involved, with each type of behavior having its own unique features and set of equations. Magnetic nanoparticles containing hundreds of spins are on the boundary between these purely classical and quantum cases, and thus they can be used to explore the transition between these two types of behavior and how the system acts at the transitory node.

Electron magnetic resonance (EMR), also known as electron spin resonance (ESR), is a convenient method to study magnetic properties and magnetization dynamics. Earlier EMR studies of nanoparticles have used a classical ferromagnetic approach, where dynamics of the magnetization was described as ferromagnetic resonance (FMR) with the Landau-Lifshits equation taking into account strong thermal fluctuations [2, 3]. However, this model hardly describes the shape and temperature dependence of the EMR line. The ferromagnetic model predicts strongly asymmetric shape with decrease in temperature while the opposite is observed in experiment. The approach also fails in describing the double-feature shape of the EMR line (see, for example, Refs. [4-10]).

An alternative model was proposed in [10], where a magnetic nanoparticle was considered as a giant spin, and the EMR signal was a sum of contributions of quantum transitions between energy levels associated with the projections of the giant spin onto the direction of the magnetic field. Although, this approach was based only on the study of γ-$Fe_2O_3$ nanoparticles that were specially fabricated and had a relatively small size (~ 5 nm diameter). It also used some phenomenological assumptions on the spectral width and fraction of free particles. To better understand the transition from quantum to classical case, more studies are necessary with various systems. The goal of this work is to get more experimental data on EMR in magnetic nanoparticles of a different type and size, and analyze the data in comparison with $\gamma$-$Fe_2O_3$.

**Experimental**

In the experiment we used the commercially available magnetite $Fe_3O_4$ ferrofluid (Amazing Magnets LLC: Ferro-Tec EFH1). The ferrofluid contains 5% magnetic solid, 10% surfactant and 85% carrier per volume. The physical picture of the particles was obtained with TEM, and size distribution was analyzed using the ImageJ software.

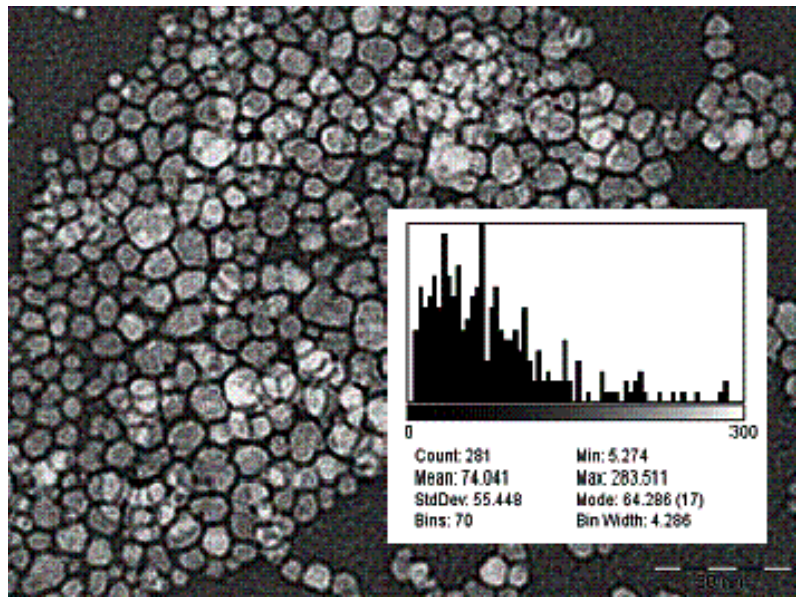

Figure 1: TEM image of $Fe_3O_4$ nanoparticles. Inset: Size distribution in terms of area ($nm^2$) of the ferrofluid nanoparticles

The ferrofluid particles are rather large as seen in Figure 1. While the size varies significantly, the median particle size falls in the range of 60-70 $nm^2$ suggesting a diameter of around 9 nm.

To prepare the experimental samples, the concentrated ferrofluid was diluted in liquid and polymer matrices. The polymer samples were fabricated by first dissolving polymer (poly-styrene co-butadiene co-methyl methacrylate) in toluene and doing the same for the concentrated ferrofluid. The dilute polymer solution was mixed with the ferrofluid solution in a beaker and sonicated for 45 minutes. The mixture was then poured on an aluminum foil and after several hours, the toluene evaporated leaving behind a solid film in which $Fe_3O_4$ particles were embedded in the polymer matrix. The liquid samples were produced by diluting concentrated $Fe_3O_4$ in toluene.

EMR studies were performed using a Bruker EMX Spectrometer operating at 9.8 GHz (X band) and modulation frequency of 100 kHz. A commercial gas flow cryostat was used in variable temperature experiments in the range 90 - 350 K. Liquid nitrogen was used for 77 K and in the Field-Freezing (FF) experiments. Concentration and temperature dependences were studied for the both liquid and polymer samples. Liquid samples also underwent FF experiments similar to those described in [4, 10]: the magnetic field was applied to the sample at room temperature, thereby partially aligning nanoparticle magnetic moments in the direction of the field. Afterwards, liquid nitrogen was poured into the cryostat, causing the ferrofluid solution to freeze with the preferable orientation of the magnetic moments. The field-frozen solution is then studied at different angle orientations with respect to the freezing field.

The data obtained in these systems are compared with the data obtained earlier with the surface modified particles of $\gamma$-$Fe_2O_3$ with a ~ 4.8 nm diameter and narrow distribution of size with the log-normal dispersion $\sigma = 0.15$. Detailed description of the samples and results of EMR studies in $\gamma$-$Fe_2O_3$ suspensions can be found in [10]. Note, that the bulk saturated magnetization of $\gamma$-$Fe_2O_3$ (~ 400 e.m.u.) is close to that of $Fe_3O_4$ (446 e.m.u), and therefore the size of a particle is expected to be one of the major factors affecting the change in behavior.

**Results**

The EMR signals in polymer and toluene matrices taken at room temperature are shown in Figure 2 for the different concentrations of the $Fe_3O_4$ ferrofluid.

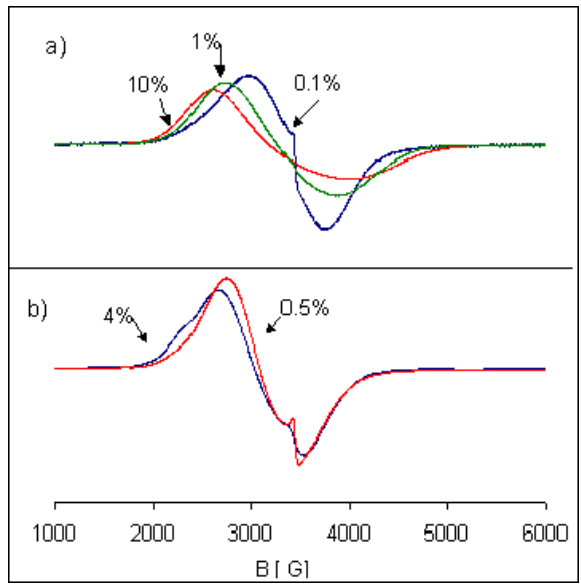

Figure 2: The EMR signal in solid (a) and liquid (b) suspensions of $Fe_3O_4$ at different concentrations as indicated in the graphs.

As one can see, in both solid and liquid matrices, the signal consists of two major components: a broad component, which shifts left and broadens with increasing concentration in the liquid sample, and a narrow component that is observed at the same field ~ 3500 G (g= 2) regardless of concentration for both liquid and solid samples. The narrow component is clearly seen in the well-diluted samples such as 0.1% (Fig 2 a) and 0.5% (Fig 2 b). Increase in the ferrrofluid concentration results in broadening of both components, and the narrow component subsequently becomes less visible until it is completely unobservable in the highly concentrated polymer samples.

The spectral shape evolution with temperature in the well-diluted polymer sample is shown in Figure 3. As one can see in Fig. 3a, decreasing temperature causes the broad component to widen, become more symmetric, and shift to lower fields. The narrow line stays at the same field, but rapidly decreases in amplitude. The spectral shape and its evolution with temperature variation are very similar to the behavior observed in $\gamma$-$Fe_2O_3$ ferrofluid [5,10], as well as in some other systems (see, for example, the data on Fe precipitates in glasses [8] and Mn clusters in LaGa(Mn)$O_3$ [11]).

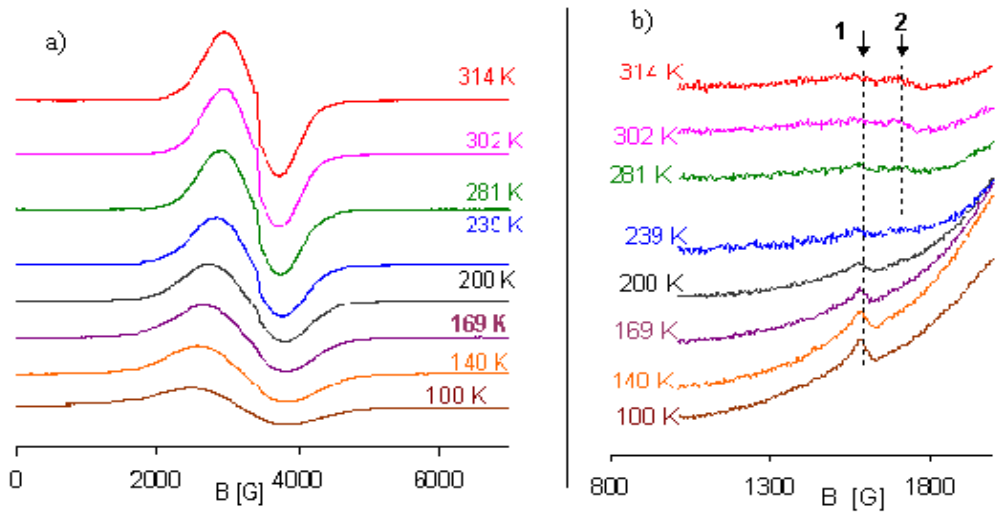

Fig. 3. a) EMR signal in the polymer sample with 0.1 wt.% of the ferrofluid at different temperatures. b) half-field range in more detail

Fig. 3 b demonstrates two additional small signals observed in the polymer sample with $Fe_3O_4$ at low fields. The first one is observed with g = 4.3 (which is typical for $Fe^{3+}$ ions) and it grows with cooling. Clearly, it has paramagnetic origin and can be ascribed to the residue of paramagnetic $Fe^{3+}$ ions in a strong crystalline field with rhombic symmetry [8, 12]. On the contrary, the second signal observed at g = 4.0 steeply decreases in the amplitude with decrease in temperature. This type of signal was reported in $\gamma$-$Fe_2O_3$ samples along with the additional signals at g= 6 and 8 in Ref. [13]. It will be further discussed later.

Figure 4 presents the results of temperature studies of the EMR spectral characteristics and FF experiments in the $Fe_3O_4$ suspension in comparison with the data obtained earlier in samples with $\gamma$-$Fe_2O_3$. According to Fig. 4a, both $Fe_3O_4$ and $\gamma$-$Fe_2O_3$ systems demonstrate broadening upon cooling, with the spectral width observed in $Fe_3O_4$ being significantly larger than that in $\gamma$-$Fe_2O_3$ samples.

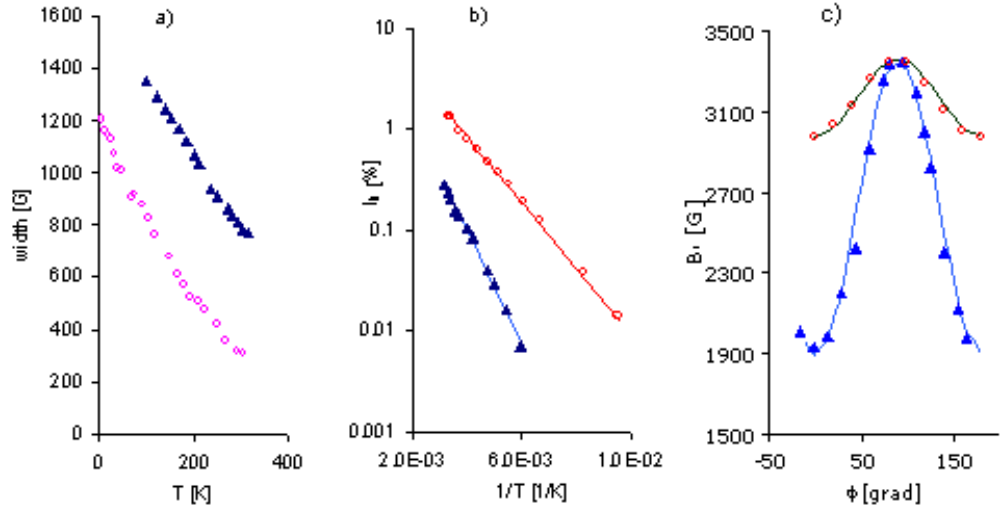

Fig. 4. a) Peak-to-peak width of the EMR signal in polymer suspensions with $Fe_3O_4$ (triangles) and $\gamma$-$Fe_2O_3$ (circles)
b) Relative intensity of narrow component (normalized to the total intensity) in polymer suspensions with $Fe_3O_4$ (triangles) and $\gamma$-$Fe_2O_3$ (circles). Solid lines correspond to the fitting by the Arrhenius law with $E/k_B$ = 1300 K and 760 K correspondingly.
c) Angular dependence of the line position of FF experiments in liquid suspensions with $Fe_3O_4$ (triangles) and $\gamma$-$Fe_2O_3$ (circles). Solid curves correspond to the $\sin^2\varphi$ dependencies.

Temperature behavior of the narrow component is shown in Fig 4 b. The relative intensity of the narrow line was estimated by the following procedure: the narrow component was fitted with the Lorentzian function, extracted from the total signal, and double integrated along with the total signal. This procedure was performed for the room temperature data, where the narrow component could be distinguished from the total signal. Then, based on the observation that the width of the narrow component does not change significantly with temperature variation, the other points in the graph of 4b were calculated using the amplitude of the derivative of the signal, similar to the procedure described in [10]. The temperature dependence of the narrow component was fitted by the Arrhenius law, $\sim\exp(-E/k_BT)$, and shown in the figure.

The results of FF experiments with the freezing field $B_{fr}$= 3500 G are shown in Figure 4c. Similar to [4, 10], the signal was shifted toward lower fields in the orientation parallel to the freezing field ($\varphi=0^o$), and toward higher fields at $90^0$. The position of the line center ($B_r$) can be fitted with a $\sin^2\varphi$ function, suggesting a uniaxial anisotropy for the system in study. The amplitude of the variation of the position $\Delta B_r = B_r(90^o) - B_r(0^0)$ was much higher in the case of $Fe_3O_4$ (1410 G) than in $\gamma$-$Fe_2O_3$ (375 G).

**Discussion**

Let us first briefly describe general characteristics of EMR in purely quantum and classical cases, and then discuss the features observed in the nanoparticles. In systems consisting of a single isolated spin or several interacting spins, the EMR (also referred to as electron paramagnetic resonance, EPR) is considered in the frames of quantum approach as resonance transitions between energy levels associated with the spin and determined by the Spin Hamiltonian as

$$H = -\mu_B \sum_i \mathbf{H} g \mathbf{S}_i + J \sum_{(i<j)} \mathbf{S}_i \mathbf{S}_j + \sum_i (D S_{z(i)}^2 + E(S_{x(i)}^2 - S_{y(i)}^2)) + H_1 , \quad (1)$$

where $\mu_B$ is the Bohr magneton, g is the g-factor, $J$ is the isotropic exchange integral, $D$ and $E$ are the anisotropy constants, and $H_1$ includes other terms such as dipole-dipole, hyperfine, and spin-lattice interactions.

Equation (1) predicts multiple energy levels for a system of several interacting spins, and a typical EPR spectrum of the spin cluster may consist of a great number of narrow lines. The dynamics of the transitions is described with a set of differential equations including corresponding set of relaxation rates which, in general, depend on the magnetic quantum numbers $m$. In solids, the longitudinal ($T_1$) and transverse ($T_2$) relaxation times may have different origin and substantially differ in magnitude, with $T_1 \gg T_2$. Magnetic phase transition does not occur in isolated spin clusters even at low temperatures.

On the other hand, systems containing statistically large number of spins coupled by strong exchange interaction (e.g. ferromagnetic) become magnetically ordered below $T_C$. Magnetization motion in such ferromagnetic domains at the external magnetic field can be described classically, with the Landau-Lifshits (LL) equation. In this case, the relaxation rates of the longitudinal and transverse magnetization components are related to each other and determined by the LL relaxation term, $R_L$. The ordinary ferromagnetic resonance (FMR) spectrum consists of a single rather broad line with approximately Lorentzian shape.

Superparamagnetic nanoparticles, including iron oxide nanoparticles considered here, are formally in a single ferromagnetic domain state, where all spins inside the particle volume are strongly coupled, giving rise to a total magnetic moment $\mu$. The size of the particle is small, and as a consequence it exhibits superparamagnetic behavior, which takes place when the magnetic energy $U_M$ is on the order of the thermal energy kT. In this regime, the direction of $\mu$ fluctuates significantly from the equilibrium position while the amplitude of $\mu$ is considered to be unchanged [2, 3].

Our experiments confirm that the EMR signal observed in superparamagnetic nanoparticles of different type and size has a very specific double-feature shape, which is most pronounced when the system in study is well diluted. Accordingly, this specific shape can be ascribed to an ensemble of non-interacting particles. Dipole-dipole interaction between the particles causes broadening of both components, which is expected and observed in highly concentrated samples. Other general features of EMR in nanoparticles were observed and confirmed as well, including the broad component becoming wider and shifting to lower fields upon cooling, temperature related behavior of the narrow component, and uniaxial character of the FF anisotropy.

In comparison with the γ-Fe$_2$O$_3$ systems, the diluted Fe$_3$O$_4$ ferrofluid demonstrates an increased width of the EMR signal, increased amplitude of the variation of the centerfield position in FF measurements, and more pronounced shift of the broad component in liquid systems (see Fig 2 b). This can be expected and readily explained in both classical or quantum considerations in terms of increased effective anisotropy due to larger size, higher irregularity of shape, and broader distribution of size and shape in Fe$_3$O$_4$ in comparison with the γ-Fe$_2$O$_3$ samples. Another factor is stronger inter-particle interactions due to closer possible distances between Fe$_3$O$_4$ particles. (The relative oxide/surfactant volume ratio was about 1: 9 in γ-Fe$_2$O$_3$ surface modified ferrofluid, and 1: 2 in the Fe$_3$O$_4$).

Let us concentrate on the features that in our opinion reflect the quantum behavior of the systems and should thus be described in framework of quantum approach. Such features include the specific shape of the EMR signal mentioned earlier, its temperature behavior (in particular, the temperature dependence of the narrow component), and the additional signals observed in the half-field range (g = 4).

The "quantization" approach discussed in [10] explains the specific shape of the EMR considering transitions between energy levels associated with the projections of the giant total spin S of a particle onto the direction of the external magnetic field **B**. According to the simplified model in the case of weak uniaxial anisotropy, the resonance field for the transition m → m+1 can be found as

$$B_{m,\theta} = B_0 + \frac{1}{g\mu_B}(2m+1)DP_2(\cos\theta), \qquad (2)$$

where *m* is the magnetic quantum number determining the projection of the total spin *S* of the particle on the direction of the magnetic field, **B**, $B_0 = \omega_0/\gamma$; $\omega_0$ is the microwave frequency, $\gamma \equiv g\mu_B/\hbar$ is the gyromagnetic ratio, $P_2(y) = (3y^2-1)/2$ is the Legendre polynomial, and *θ* is the angle which **B** makes with the anisotropy axis.

In this model [10], the narrow line is ascribed to the transitions between the central (excited) energy levels having small magnetic quantum numbers |m|. The corresponding resonance fields are weakly affected by anisotropy terms, see Eq. (2). Therefore, this part of the spectrum is not broadened by random distribution of the anisotropy axes. In contrast, the broad EMR component is caused by the sum of contributions from transitions with large values of |m|, whose positions are shifted from $B_0$ toward lower or higher fields depending on the orientation of the anisotropy axes. Averaging over all possible orientations results in the spreading of the satellites to a "powder" pattern with the width determined by the anisotropy parameters. To better fit the experimental data, some additional phenomenological assumptions were accepted [10] about the width of the line and fraction of free (not aggregated) particles, which implicitly determine the relative number of the transitions contributing to the narrow component.

Comparing the intensities of the narrow component and the total signal in Fe$_3$O$_4$ and γ-Fe$_2$O$_3$, one can see that the relative intensity of the narrow component $I_N/I_0$ is significantly lower in Fe$_3$O$_4$ than in γ-Fe$_2$O$_3$ throughout the whole temperature range (Fig. 4 b). At room temperature, the ratio $I_N / I_0 \sim 0.2\%$ was obtained for Fe$_3$O$_4$ systems, in contrast with ~ 1.1% in γ-Fe$_2$O$_3$. The difference becomes more pronounced at lower temperatures. If the fitting traces of Fig 4b were extrapolated to the high temperature range, they are not crossing below ferromagnetic transition temperature. Thus, the intensity of the narrow component is always much lower, and it vanishes much more rapidly with cooling in larger-sized particles. Apparently no such narrow feature is expected from the large systems in the classical case. Note that the decreasing of the narrow component with an increase in particle size was also reported in Ref. [5].

The "quantization" model explains temperature behavior of the components as well, linking them to populations of the energy levels contributing to the particular component. With decrease in temperature, growing population of lower levels contributes more in the broad component, making it higher in intensity and broader. Meanwhile, the central transitions become depopulated, causing a rapid drop in intensity of the narrow component, $I_N$, as was observed in experiment. At $U_M > kT$, the population difference between the central neighboring levels is approximately proportional to $\exp(-U_M/kT)$, where $U_M = MVB$, V being the particle volume and M the saturated magnetization. As one can see in Fig. 4b, in both $Fe_3O_4$ and $\gamma$-$Fe_2O_3$, the temperature dependence of the narrow line intensity does follow an exponential law. In $Fe_3O_4$ the slope is higher than in $\gamma$-$Fe_2O_3$. In the ideal case, nearly 8–fold increase of the activation energy can be expected for the 9 nm-size particles $Fe_3O_4$ in comparison to 4.8 nm size $\gamma$-$Fe_2O_3$ particles. However, only twofold increase of the slope is observed (Fig. 4b). This can be explained by significantly broader distribution P(V) of the particle size in the magnetite samples (Fig. 1). Indeed, the contribution from the central (excited) energy levels of the particles having a volume V is proportional to $V \cdot P(V) \cdot \exp(-MVB/kT)$. Thus, due to the Arrhenius exponent, the smaller particles can produce larger signals. As a result, the "effective" particle diameter estimated from the temperature dependence of the narrow component may appear to be significantly lower than its mean value. In any case, the data clearly demonstrate the general tendency: an increase of the slope with the increase of the particle size.

Our experiments confirm the presence of another interesting feature in EMR of the nanoparticles: the signals observed at $B_0/n$. The signals with n = 2, 3, and 4 were reported earlier in $\gamma$-$Fe_2O_3$ [13]. Current studies demonstrate such signals (at least, with n = 2) in different particles. We believe these features have quantum origin related to multiple quantum transitions. Such transitions are well known for the paramagnetic resonance of small spin systems and explained with an admixing of the adjacent quantum states due to the anisotropy terms or dipolar interactions between paramagnetic centers. The theory [13] predicts nearly constant value of the relative signal intensity at high enough temperatures and proportionality to (kT/MVB) at low temperatures. However, much more rapid decrease with cooling is observed in the $\gamma$-$Fe_2O_3$ superparamagnetic nanoparticles [13].

The intensity of the half-range signal in $Fe_3O_4$ is very low. It can be estimated to be on the order of $8 \cdot 10^{-4}$ of the total EMR intensity while in the $\gamma$-$Fe_2O_3$ system (that has a smaller particle size) this ratio is higher and at about $3 \cdot 10^{-3}$ of the total EMR intensity. This tendency is consistent with the quantum description [13].

In summary, magnetic nanoparticles demonstrate special features in EMR behavior, such as the temperature-dependent narrow spectral component and multiple-quantum transitions, which are common for small quantum systems. With an increase of the system size, the relative intensity of these features significantly decreases, suggesting a route of a gradual evolution from a purely quantum case of a small spin system to a classical behavior of statistically large systems, where such quantum features are not expected and observed.

The quantization model [10] based on the consideration of the ground spin multiplet with well-distinguished magnetic sublevels explains the existence and some general tendencies observed in the spin systems of the intermediate size but is not sufficient to describe the whole picture. More experimental and theoretical work is needed to describe both quantum features and gradual transition to the classical dynamics in the magnetic nanoparticles and spin clusters.

The work was partly supported by National Science Foundation (NSF) CREST Project HRD-9805059, and NSF PREM Grant # DMR-0611430. One of the authors (V.A.A.) acknowledges the support of Russian Foundation for Basic Research (Project 05-02-16371) and Russian Academy of Sciences (Project P-03-2-24).